# Characterizing Self-Heating Dynamics Using Cyclostationary Measurements

Shin, SangHoon[a)], Muhammad Masuduzzaman[b)], Muhammad Ashraful Alam[c)]

School of ECE, Purdue University, West Lafayette, Indiana 47907, USA

## Abstract

Self-heating in surrounding gate transistors can degrade its on-current performance and reduce lifetime. If a transistor heats/cools with time-constants less than the inverse of the operating frequency, a predictable, frequency-independent performance is expected; if not, the signal pattern must be optimized for highest performance. Typically, time-constants are measured by expensive, ultra-fast instruments with high temporal resolution. Instead, here we demonstrate an alternate, inexpensive, cyclostationary measurement technique to characterize self-heating (and cooling) with sub-microsecond resolution. The results are independently confirmed by direct imaging of the transient heating/cooling of the channel temperature by the thermoreflectance (TR) method. A routine use of the proposed technique will help improve the surrounding gate transistor design and shorten the design cycle.

[a)] Electronic mail: shin136@purdue.edu

[b)] Electronic mail: masudworld@gmail.com

[c)] Electronic mail: alam@purdue.edu

S. H. Shin and M. Masuduzzaman contributed equally to this study

## I. INTRODUCTION

The surround gate geometry of FINFET and Gate-all-around (GAA) transistors improve short channel performance at the expense of increased self-heating [1]-[5]. The restriction of heat conduction pathways is reflected in fast (~$\mu$sec scale) self-heating of the channel and reduced on-current. The transient heating and cooling characteristics of such transistors must be known independently, so that one can (i) predict circuit performance and reliabilities at high frequencies and for a variety of non-periodic data streams, and (ii) interpret its features to identify/ameliorate the bottlenecks of heat dissipation pathways.

Among the existing characterization methodologies, some measure the internal channel temperature ($\theta_C$) electrically, while others focus on measuring of the surface temperature ($\theta_S$) optically. Each group contains fast and slow variants, with the corresponding merits and limitations. For example, the transient on-current (and $\theta_C$) can be measured with an ultra-fast equipment [4],[5], however, the setup requires specialized circuitry, such as high-bandwidth current-to-voltage converters, etc. A slower spectroscopic method involves measuring the output conductance of a transistor at varying frequencies [6]; the approach determines the net time constant, but the time-constants for transient heating and cooling cannot be determined independently.

Here, we present a simple electrical characterization method that can measure the transient heating and cooling processes separately with a sub-microsecond temporal resolution. The approach relies on DC current, and thus, ultra-fast equipment is not needed. In the following, we discuss the theory and experimental validation of the proposed technique.

## II. SELF-HEATING CHARACTERIZATION

Once a transistor is turned on, the power dissipation in the channel $P_D(t)$ increases the channel temperature $\theta(t)$, with respect to its nominal off-state value, $\theta_0$. At steady state, the temperature ($\theta_{SS}$) is defined by the balance of the heat generation vs. heat dissipation through various thermal pathways (such as substrate, gate metals, source-drain). The time to reach $\theta_{SS}$ (characterized by the heating time constant, $\tau_H$), as well as the magnitude of the steady state temperature rise ($\Delta\theta_{SS} = \theta_{SS} - \theta_0$) depend on the power dissipation and the thermal resistance ($R_{TH}$) of the heat conduction channels, ($\Delta\theta_{SS} \propto P_D R_{TH}$). Similarly, when the transistor is turned off, power dissipation is reduced and the channel temperature reduces to $\theta_0$, with a characteristic cooling transient time constant $\tau_L$. Obviously, one can use different electrical techniques to measure $\theta_{SS}$, such as AC conductance, 4T-gate resistance [4]; however, in order to know the characteristic time constants for heating and cooling processes, in the following, we will dynamically stabilize the channel temperature at different intermediate levels between $\theta_0$ and $\theta_{SS}$ by applying appropriate gate pulses. The corresponding $I_{DS}$ as a function of pulse timing will define $\tau_L$ and $\tau_H$.

**Characterization technique**: The proposed method is analogous to a traditional charge pumping (CP) technique [7], where the source is grounded and a square wave is applied at the gate; however, unlike CP, the drain is connected to $V_D$ and the body contact is not required, see Fig. 1(a). The gate pulse turns the device on ($V_{G,HI}$) and off ($V_{G,LO}$) for duration $T_H$ and $T_L$, respectively, so that the duty cycle, $d \equiv \frac{T_H}{T_H+T_L}$. The measured drain current is defined by the average on ($\overline{I_{ON}}$) and off currents ($\overline{I_{OFF}}$), as

$$I_{MEAS} \equiv d\overline{I_{ON}} + (1-d)\overline{I_{OFF}}. \quad (1)$$

Assuming that $\overline{I_{OFF}}/\overline{I_{ON}} \ll d$, the average on-current during $T_H$ from (1) is given by $\overline{I_{ON}} \sim I_{MEAS}/d$. To characterize the heating transient, *we keep $T_L$ fixed, but increase $T_H$ gradually*. For $T_H = 0$, the channel temperature is equal to $\theta_0$. As $T_H$ is increased, the device is being heated for longer duration ($\theta > \theta_0$). Once $T_H > \tau_H$, the temperature begins to saturate ($\theta_{SS}$). Thus, $\overline{I_{ON}}(T_H)$ gives a temporal measure of $I_{DS}(t)$ during self-heating. Similarly, the cooling transient is characterized by *keeping $T_H$ fixed, but gradually increasing $T_L$*. The longer off-state (increased $T_L$) leads to more cooling, which in turn reduces the average temperature for bias $V_{G,HI}$ and increases $\overline{I_{ON}}$ as a function of $T_L$ (until $T_L$ is large enough to saturate the cooling process). Thus, one can determine $I_{DS}(t)$ during the cooling process by monitoring $\overline{I_{ON}}(T_L)$ (with fixed $T_H$), see Fig. 1(b-c).

**A model of heating and cooling transients**: Although the time-constants will be obtained directly from the measurement data independent of any theoretical model, the following simple model rationalizes the approach. Specifically, to show theoretically that $I_{ON}$ responds to the $T_H$ and $T_L$ variation, we use a system described by the first order rate equation below,

$$V\rho C_p \frac{d\theta}{dt} = P_D + \ A \times h\ (\theta - \theta_0) \tag{2}$$

where the volume $V$, material density $\rho$, specific heat capacity $C_p$, input power $P_D = I_D \times V_D$ , area $A$, and heat transfer coefficient $h$ define the thermal system and is characterized by the thermal time constant, $\tau \equiv \rho C_p/sh$, where $s \equiv A/V$ is the surface-to-volume ratio. Modern transistors have complex, multilayer stacks with complicated heat conduction pathways; in this context, $\tau$ should be interpreted as an effective time constant.

When a periodic $P_D$ (with duty cycle $d$) is applied by turning on and off the gate voltage ($V_G$) and a steady state is reached, the transistor temperature would cycle between $\theta_{init,L}$ and $\theta_{init,H}$, at the beginning and the end of each pulse. These cyclostationary temperatures can be obtained by requiring that $\theta(t) = \theta(t + T_H + T_L)$. In other words, the temperature at the beginning of $V_{G,HI}$ ($\theta_{init,H}$) should be the same as that at the end of $V_{G,LO}$. The system can be described by the following equation,

$$\begin{bmatrix} 1 & -e^{-\frac{T_H}{\tau_H}} \\ -e^{-\frac{T_L}{\tau_L}} & 1 \end{bmatrix} \begin{bmatrix} \theta_{init,L} \\ \theta_{init,H} \end{bmatrix} = \begin{bmatrix} \theta_{SS}\left(1 - e^{-\frac{T_H}{\tau_H}}\right) \\ \theta_0\left(1 - e^{-\frac{T_L}{\tau_L}}\right) \end{bmatrix} \tag{3}$$

Since $\theta_{SS} \propto P_D \propto I_{DS}V_{DS}$, once Eq. (3) is solved for $\theta_{init,j}$, $j = H, L$; $\theta(t)$ can be determined for all bias conditions. Note that, the relative magnitude of $T_H$ and $T_L$ determines the two boundary conditions for $\theta(t)$and hence the average temperature during on-state ($\theta_{avg,H}$), which in turn, determines $\overline{I_{ON}}$ (Recall that $I_{ON} \propto \mu_{ph} \sim \theta^{-3/2}$[8], where $\mu_{ph}$is the mobility due to lattice scattering). Indeed, numerical simulation confirms that $\theta_{avg,H}$ can be controlled by varying $T_H$ and $T_L$, as explained in detail in the supplementary material; For example, the $T_H$ sweep (fixed $T_L$) results in an increase in $\theta_{avg,H}$ and a drop in $\overline{I_{ON}}$, approximately with a time constant $\tau_H$ (and vice versa), as predicted by our characterization algorithm (see Fig. 1(e)).

## III. EXPERIMENTAL RESULTS AND VERIFICATION

**Self-heating in gate-all-around (GAA) transistors:** For experimental demonstration, we use GAA InGaAs n-MOSFETs (channel length = 60nm, 4 parallel nanowires with cross section of 30×35nm$^2$ each, EOT = 1.7nm) [9], because significant self-heating is anticipated. The gate stack consists of $Al_2O_3$ covered by the gate metal layer (WN). Our previous experiments show that even a moderate $V_{DS} = 1\ V$ leads to significant self-heating. The gate pulse with $V_H > V_{th}$(threshold voltage) and $V_L < V_{th}$ turns the transistor on and off, respectively. It is preferable to choose $V_H$ and $V_L$ so that highest on-off ratio is achieved, and the condition $\overline{I_{ON}}/\overline{I_{OFF}} \gg d^{-1}$ is satisfied even when $d$ is small.

Fig. 1(b) shows the experimental $\overline{I_{ON}}$ (extracted as $I_{MEAS}/d$) using the setup of Fig. 1(a) for varying $T_H$ and $T_L$. In principle, one can obtain the thermal time-constants by analyzing the current along any vertical and horizontal trajectory in the $T_H$ and $T_L$ plane (see Fig. 1(b). To ensure the condition $\overline{I_{OFF}}/\overline{I_{ON}} \ll d$ is satisfied, however, we take the heating transient data for a lower fixed value of $T_L$ (Fig. 1(b), red arrow) and the cooling transient data for a higher fixed value $T_H$ (Fig. 1(b), green arrow). As shown in Fig. 1(c), the heating and cooling transients have time constants on the order of microsecond, consistent with other reports of GAA systems as well as SOI devices [1]-[5]. The asymmetry in time constants $\tau_H = 581$ns and $\tau_L = 695$ns suggests that the system is nonlinear. Indeed, this asymmetry may have important implications in determining the average temperature of a transistor operating at a GHz frequency [10]. For reference, $R_{TH}$ calculated approximately $R_{TH} = 657.98\ \mathrm{K/mW}$, consistent with other reports in the literature [see Supplementary Materials, Sec. 4].

**Electrical Validation of the Characterization Approach**: Since $\theta - \theta_0 \sim P_D \times R_{TH}$, one can validate the approach described above in two ways, namely, (i) by demonstrating that $\theta \rightarrow \theta_0$ as $P_D \rightarrow 0$ (e.g. reducing drain voltage for GAA devices) or, (ii) by setting $R_{TH} \rightarrow 0$, using a planar device, because it has relatively few thermal bottlenecks [6]. For the first case, if we use $V_D = 50\ mV$ instead of $1\ V$, $P_D$ is reduced by two orders of magnitude ($V_D$ and $I_{DS}$ reduce by a factor of 20 and 15, respectively, compared to $V_D = 1V$); therefore, the self-heating is negligible at low $V_{DS}$. Fig. 1(d) shows the results of the measurement at $V_{DS} = 50mV$. Since self-heating is absent, we attribute the 7% loss in $\Delta\ I_{ON}$ to the shift in the threshold voltage due to charge trapping. Assuming that that the trapping component is independent of $V_D$, we can decouple the effect of self-heating by subtracting the non-heated characteristics (Fig. 1(d) for $V_{DS} = 50mV$) from that of the self-heated device (Fig. 1(c) for $V_{DS} = 1V$), see supplementary material, Sec. 3. Additionally, we carefully selected bias conditions of $V_G = 0.8\ V$ and $V_D = 1V$ for the period of $T_H$ to avoid contamination from hot carrier injection (HCI) and PBTI degradation of $I_{ON}$. Specifically, for a stress duration of $10^4 sec$, PBTI ($V_{GS} = 1.8V$ and $V_{DS} = 0V$) and HCI ($V_{GS} = V_{DS} = 2V$) degrades $\Delta I_{ON}$ is about 5% and 35% respectively. In contrast, the entire cyclostationary dataset in Fig. 1(b) collected in less than 5 minutes and each set of measurements takes $1sec$ with $3sec$ interval, at substantially reduced voltage; therefore, we do not expect significant contributions from NBTI, HCI, or PBTI effects [3], [11]. Finally, as a reference, we measured the self-heating of choose a planer device with Si substrate. Here, self-heating should be negligible even for high drain bias, because heat can easily escape through the substrate (i.e. low $R_{TH}$). The proposed measurement technique confirms this assertion: Fig. 1(f) shows that even at $V_D = 1V$, the change in $I_{ON}$ during the heating and cooling processes are negligible, implying that the self-heating is insignificant.

**Optical Validation of the Characterization Approach**: The self-heating and the thermal time constants can also be measured directly by thermoreflectance (TR) imaging [12], which provides an independent verification of the results in Fig. 1(c). In this technique, the gate surface temperature ($\theta_S$) of a device is imaged by an ultra-fast measurement technique with high spatial resolution (Fig. 2(a)); the detailed measurement process is described in [12],[13].

Fig. 2(b) (inset) shows images taken at different time instants of heating process. Fig. 2(b) shows that the heating transients from the optical TR method and the on-current transient obtained from the proposed electrical technique agree remarkably well, both characterized by similar time constant $\tau_H \sim 1\ \mu s$. A 3D thermal modeling interprets the time constant with self-heating of the source and drain contact-pads of the transistors [14].

## IV. CONCLUSIONS

We have demonstrated an experimental technique to characterize the self-heating transients for heating and cooling processes independently, with a temporal resolution of $100\ ns$. The technique is based on electrical measurement and can directly reflect the temperature rise within the channel. The technique is theoretically simulated for different gate pulses. The validity of the approach is confirmed with several independent experiments. In agreement with other reports in the literature, our results show that the heating and the cooling transients have a characteristic time constants in the microsecond range, but they need not be symmetric. This technique can be used to characterize the self-heating for process optimization as well as to predict the temperature at the high operating frequency by precisely characterizing the heating and cooling dynamics.

## IV. ACKNOWLEDGEMENT

We would like to thank Nathan Conrad and Prof. Peide Ye for samples and useful discussion, Prof. Ali Shakouri for Thermoreflectance imaging facility, and Birck Nanotechnology Center at Purdue University for experimental facilities for this work.

**Figures:**

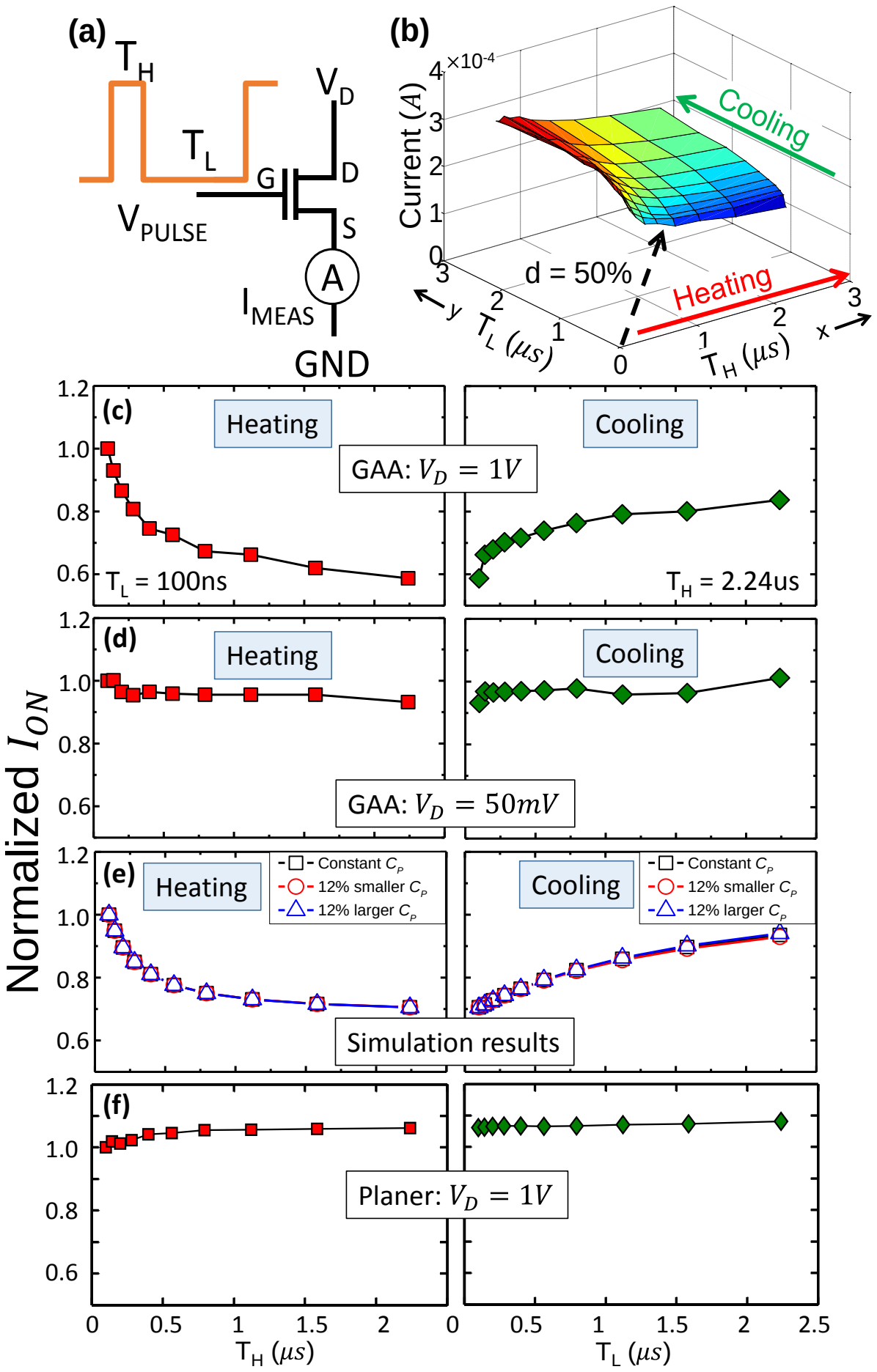

Figure 1: (a) A schematic diagram for the experimental setup. (b) Experimental data for $I_{ON}$ as a function of $T_H$ and $T_L$ with fixed $T_L = 100ns$ and $T_L = 2.24us$ respectively, and the particular sweep directions heating (red) and cooling (green) measurements as compared to a spectroscopic measurements with a 50% duty cycle (black-dotted). (c) The data along two particular lines in (b) are shown representing the heating and cooling transients at high $V_D$ condition. (d) The same data with low $V_D = 50mV$ for the GAA transistor. (e) Simulated ION shows similar transient behavior with $T_H$ and $T_L$ variation and different $C_p$ as in (c). (f) The heating transient data of $I_{ON}$ for planar Si devices ($R_{TH} \rightarrow 0$).

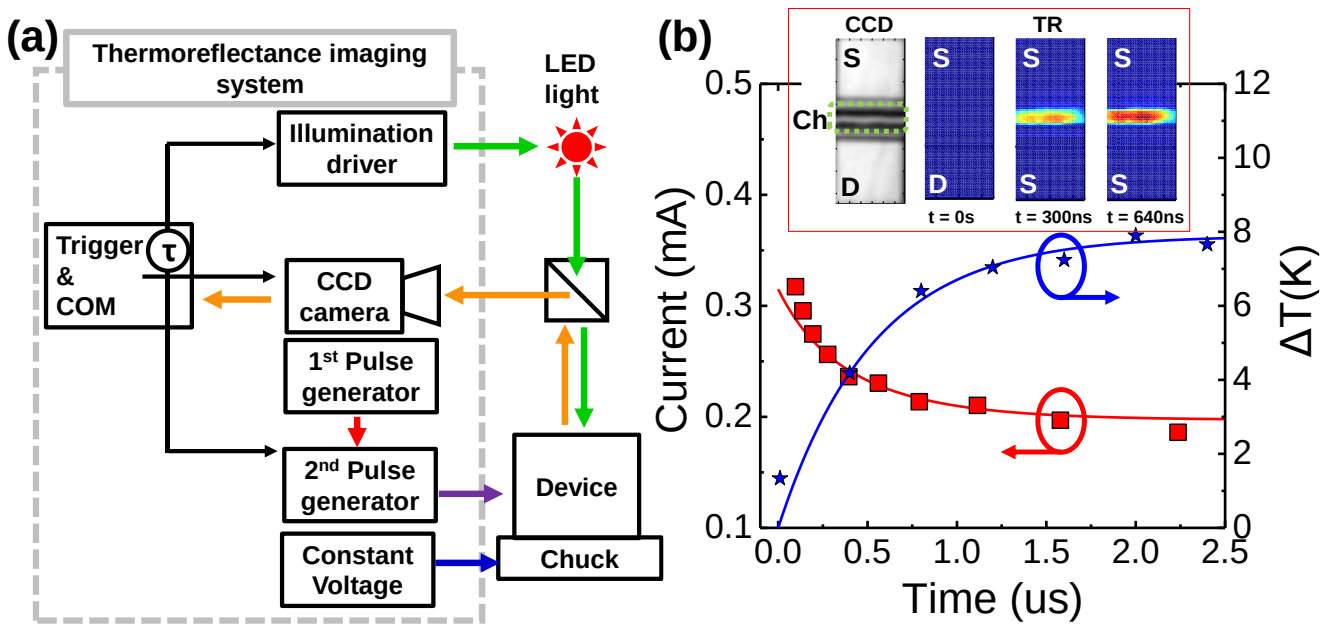

Figure. 2: (a) The experimental setup for a thermoreflectance measurement (left). (b) The optical as well as the TR images are shown for a few time instants (inset).The temperature rise as measured from the TR method (⋆) compares well with the transient current decay (■) measured from the proposed electrical characterization technique (right). Solid lines for each symbols are current vs. mobility for phonon scattering relationship ($I_{ON}(t) \propto \mu_{ph} \sim \theta(t)^{-3/2}$) based fitting.